\documentclass[a4paper,prb,twocolumn]{revtex4}
\pdfoutput=1
\usepackage{amsfonts}
\usepackage{graphicx}
\usepackage{color}
\usepackage{amsmath}
\usepackage{amssymb}
\usepackage{latexsym}
\usepackage{psfrag}

\begin{document}
\title{Self-consistent Hartree theory of the role of electron-electron interactions in the electronic structure and conductance quantization of graphene nanoconstrictions}
\author{S. Ihnatsenka}
\affiliation{Department of Physics, Simon Fraser University, Burnaby, British Columbia, Canada V5A 1S6}
\author{G. Kirczenow}
\affiliation{Department of Physics, Simon Fraser University, Burnaby, British Columbia, Canada V5A 1S6}

\begin{abstract}
We present self-consistent calculations of electron transport in graphene nanoconstrictions within the Hartree approximation. We consider suspended armchair ribbons with V-shaped constrictions having perfect armchair or zigzag edges as well as mesoscopically smooth but atomically stepped constrictions with cosine profiles. Our calculations are based on a tight-binding model of the graphene and account for electron-electron interactions in both the constriction and the semi-infinite leads explicitly. We find that electron interactions result in (i) Electrons accumulating along edges of the uniform ribbon and along zigzag and cosine constriction edges but not along armchair constriction edges. (ii) The first subband showing almost perfect transmittance due to localization at the uniform graphene boundary except at low energies for the cases of zigzag and cosine constrictions where Bloch stop-bands form in related periodic structures. (iii) The second subband being almost perfectly blocked by the constriction. (iv) Electron interactions favor intra-subband scattering while the non-interacting electron theory predicts predominance of inter-subband scattering. (v) Conductance quantization for the first few conductance steps being more pronounced for armchair constrictions but less so for zigzag constrictions. (vi) A much more prominent $2e^2/h$ conductance plateau for the cosine constriction than is found in the absence of electron interactions. Possible implications for recent experiments are briefly discussed.
\end{abstract}

\maketitle

\section{Introduction}

Graphene nanoribbons are strips of graphene several nanometers wide and of arbitrary length.
Their unique electronic structure and transport properties that arise from the linear, massless
Dirac-like spectrum of the underlying honeycomb lattice of graphene are attracting a great deal
of interest at the present time.\cite{review} In experimental studies of graphene nanoribbons
the electronic charge density in the ribbon is usually varied by the application of a variable 
voltage to a gate electrode located near the ribbon.\cite{Han07, Chen07, Lin08,Li08sonif,Wang08, 
Molitor09, Stampfer09,Jiao09, Todd09} If, due to the application of the gate voltage, there
is a net charge on the ribbon, the electronic charge density in the ribbon ceases to be uniform,
and there is a strong redistribution of the charge towards the edges of the ribbon. This charge
redistribution and its effects on the electronic structure and transport have
been examined theoretically by several authors for ribbons of uniform width separated from
the gate electrode by a dielectric film.\cite{FR07, Silvestrov08, Shylau09, others_ee} 
It has been predicted that the charge
redistribution results in a $1/\sqrt{x}$ charge singularity\cite{Silvestrov08} at the edge of the ribbon (here $x$ is the coordinate normal to the ribbon boundary) 
and modification of the electron dispersion relation.\cite{Shylau09}
Electron transport in graphene nanoribbons with constrictions (GNCs) is also
attracting theoretical\cite{MunozRojas06, MunozRojas08, Guimaraes12, constriction12}
 and experimental\cite{Tombros11} attention at the present time.
However, the effects of charge redistribution in GNCs have not as yet been discussed in the literature. 
In this paper we explore this topic theoretically. 

The present theory accounts for the effects of the Coulomb repulsion that gives rise to the charge redistribution within the self-consistent Hartree approximation that has been used previously to study charge redistribution in uniform ribbons.\cite{FR07, Silvestrov08, Shylau09, others_ee} 
We consider infinite ribbons with armchair edges and V-shaped or cosine shaped constrictions and treat the effects of electron-electron interactions at the Hartree level throughout these entire structures. The ribbons that we consider are suspended above a dielectric layer which covers a gate electrode as in a recent experiment.\cite{Tombros11} We treat the effects of the dielectric and gate within an image charge model. We describe the charged nanoribbon by fixing the chemical potential, and obtaining a self-consistent solution where all electronic states with lower energies than the Fermi energy are filled. From this calculation we obtain the Hartree electronic energy bands and charge densities, and compute the conductance. 
Our model considers only $p_z$-orbitals of the graphene within the tight-binding approximation. The structures studied below include more than 10000 carbon atoms inside the computational area.

We find that strong charge accumulation along the constriction boundary occurs or does not occur depending on the type of constriction that is considered. As more electron subbands become populated the lowest subbands gradually localize near the graphene boundary. We find that, in the Hartree model, electrons in the first subband of the ribbon are transmitted almost perfectly through the constriction with little intersubband scattering, except at the small Fermi energies where they can be resonantly reflected by a constiction having zizgag edges. By contrast, in the non-interacting electron model strong intersubband scattering occurs at the constriction. Depending on the type of constriction conductance quantization is predicted to be more or less pronounced in the Hartree approximation than in the non-interacting approximation, a finding that may be relevant to recent experimental observations. 

The remainder of this paper is organized as follows: Our model and method of solution are described in Section \ref{Model}. Our numerical results are presented in Section \ref{Results}. We summarize our main conclusions in Section \ref{Summary}.

\section{Model}
\label{Model}
\begin{figure}[b]
\includegraphics[scale=1.0]{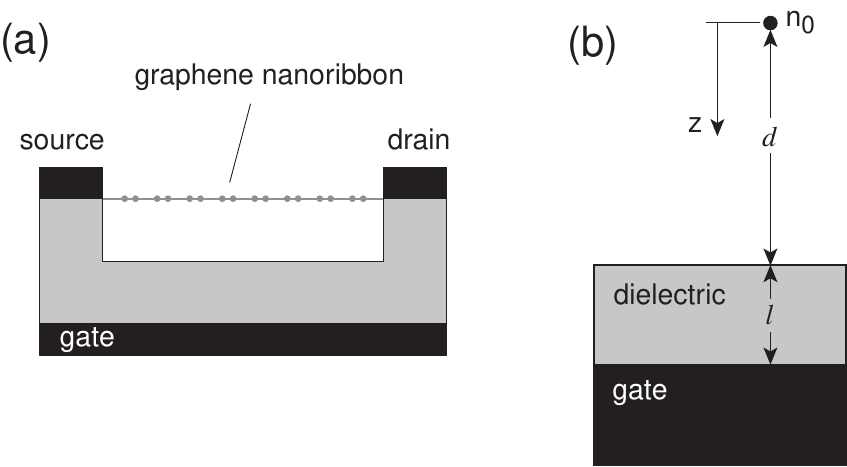}
\caption{(a) Representative device with freestanding graphene nanoribbon. (b) Image charge model.
}
\label{fig:1}
\end{figure}

We consider a graphene nanoribbon (GN) suspended in a way similar to that in the experimental setup of Tombros etal.\cite{Tombros11} The GN is separated from the back gate by layers of a dielectric and air, see Fig. 1(a). For the dielectric material we choose SiO$_2$ with relative permittivity $\epsilon=3.9$. The NR is attached at its two ends to semiinfinite leads represented by ideal ribbons having the same width $W$ as the NR. Four different types of devices are considered in the following: an ideal uniform ribbon, and ribbons having V-shaped armchair and zigzag and constrictions or constrictions with cosine profiles imposed on one side.  We disregard any defects other than the atomic steps at the boundary of the cosine-shaped constriction in the present study. The host configuration is taken as armchair as are the edge configurations of the semiinfinite leads.

As representative devices we consider GNs of width $W=10$ nm and length $L=27$ nm, Fig. 1(a). There are 92 carbon atoms in the cross-section making the ribbon semiconducting.  The constrictions, if imposed, are V-shaped or cosine-shaped trenches 5 nm deep inside the ribbon thus leaving half of the width $W$ of the ribbon in the narrowest part of the constriction for electron propagation. The length $L$ is taken long enough to include part of the leads near the constriction and therefore to treat the leads accurately. Because for the V-shaped constrictions we consider atomically ideal boundaries, the apex angles of these constrictions are 60 and 120 degrees for armchair and zigzag cases, respectively. The SiO$_2$ and air layers are both 50 nm thick. Thus the back gate is 100 nm from the nanoribbon. We performed simulations for different ribbon widths for these constriction geometries and all of the results showed similar features.

The system shown in Fig. 1 is described by the Hamiltonian 
\begin{equation}
  H = \sum_i V^H_i\;a_i^{\dag }a_i - \sum_{\left\langle i,j\right\rangle} t_{ij}\left( a_i^{\dag}a_j + h.c. \right),
  \label{eq:hamiltonian}
\end{equation}
where $t_{ij}=t=2.7$ eV is the matrix element between nearest-neighbor atoms; $V^H_i$ is the Hartree potential at atom $i$ which results from the Coulomb interaction with the uncompensated charge density $-en$ in the system (including the image charges). In coordinate space the Hartree potential can be written as
\begin{equation}
  V^H(\mathbf{r})=\frac{e^{2}}{4\pi \varepsilon _{0}\varepsilon}\int d\mathbf{r}\,^{\prime } \sum_k \frac{n_k(\mathbf{r}^{\prime})}{\sqrt{|\mathbf{r}-\mathbf{r}^{\prime}|^{2}+b_k^{2}}}  ,
  \label{eq:V_H}
\end{equation}
where $-en_k(\mathbf{r}^{\prime})$ is the $k^{th}$ electron or image charge placed at distance $b_k$ from the graphene layer. The image charges included in the model keep the back-gate electrode at zero potential.\cite{Shylau09, Silvestrov08, FR07, Peres09} The potential due to a charge density $-en_0$ located a distance $d$ above a dielectric (with dielectric constant $\varepsilon$ and thickness $l$) that is over a metal gate, as shown in Fig. \ref{fig:1}(b), can be described by an infinite number of image charge densities. The first few image charge densities and their $z$-coordinates (measured from the position of the electron charge) and also the electron charge density itself are given in Table \ref{tab:1}. The first row ($k=0$) describes the direct Coulomb interaction between electrons in the graphene layer. Because the contributions from the image charges  decrease rapidly as $k$ grows and in order to facilitate computation the results presented below were obtained keeping only the $k=0,1,2,3$ terms. 

\begin{table}
	\centering
		\begin{tabular}{c | c | c}
		\hline
$k$ & $z$ coordinate, $b_k$     & charge density, $-en_k$ \\		
\hline	
0 & 0         & $-en_0$  \\
1 & $2d$      & $-e\frac{1-\varepsilon}{1+\varepsilon}n_0$   \\
2 & $2(d+l)$  & $e\frac{4\varepsilon}{(1+\varepsilon)^2}n_0$ \\
3 & $2(d+2l)$ & $e\frac{4\varepsilon}{(1+\varepsilon)^2}\frac{1-\varepsilon}{1+\varepsilon}n_0$  \\
4 & $2(d+3l)$ & $e\frac{4\varepsilon}{(1+\varepsilon)^2}\left(\frac{1-\varepsilon}{1+\varepsilon}\right)^2n_0$  \\
... & ... & ... \\
\hline
		\end{tabular}
	\caption{Coordinates and electron and image charge densities for the model shown in Fig. \ref{fig:1}(b). $k=0$ refers to
	the electron density $n_0$ that gives rise to the image charge densities for which $k>0$. $d$ is the distance between
	the graphene and the dielectric and $l$ is the thickness of the dielectric.}
	\label{tab:1}
\end{table}	

The integration in \eqref{eq:V_H} was performed over the whole device including the semiinfinite leads. In order to include electron-electron interactions over the whole system, we partition the system into three parts, the internal computational region and two semi-infinite leads.\cite{qpc,opendot} The internal region incorporates not only the constriction but also segments of uniform ribbon on both sides of it, including part of the leads. The semi-infinite leads themselves begin far enough from the constriction to ensure that the total self-consistent potential and the electron density do not change appreciably along the leads, i.e., the electron density and the potential in the leads are not affected by the internal region. Thus the leads can be considered as uniform graphene ribbons. 

Starting from the Hamiltonian Eq. \eqref{eq:hamiltonian}, we evaluate the Green's function numerically using the technique described by Xu et al.\cite{Xu08}. The Green's function in the real-space representation, $G(\mathbf{r},\mathbf{r})$, provides information about the local density of states at site $\mathbf{r}$,
\begin{equation}
  LDOS(\mathbf{r},E)=-\frac{2}{\pi S}\Im \left[\mathcal{G}(\mathbf{r},\mathbf{r},E)\right],
  \label{LDOS}
\end{equation}
where factor 2 takes account of the spin degeneracy and $S$ is the area corresponding to one carbon atom. The density of states is $DOS(E)=\int d\mathbf{r} \;LDOS(\mathbf{r},E)$. The LDOS can be used to calculate the electronic density $n(\mathbf{r})$ at site $\mathbf{r}$
\begin{equation}
  n(\mathbf{r})=\int_{V_c}^{\infty} dE\,LDOS(\mathbf{r},E)\,f(E-E_F),
  \label{density}
\end{equation}
where $E_F$ is Fermi energy and $f$ is the Fermi-Dirac distribution function. All the calculations reported in this paper correspond to the temperature $T = 10$ K. A zero temperature version of Eq. \ref{density} has been used previously in Ref. \onlinecite{Shylau09}. In general at non-zero temperatures the lower limit of integration in Eq. \ref{density} is $-\infty$. However, in the present work the LDOS is zero in a range of energies of width much larger than $kT$ below the charge neutrality point $V_c$ and therefore Eq. \ref{density} is a good approximation. The position of the charge neutrality point $V_c$ at a given Fermi energy is determined numerically from solution of the Schrodinger equation 
\begin{equation}
  H\Psi=E(k)\Psi,
  \label{eq:Schrodinger}
\end{equation}
with $H$ being the Hamiltonian \eqref{eq:hamiltonian} and the wave function $\Psi$ obeying the Bloch theorem
\begin{equation}
  \Psi_{m+M} = e^{ikM} \Psi_m,
  \label{eq:Bloch}
\end{equation}
where $k$ is the Bloch wave vector and $\Psi_m$ is the Bloch wave function at coordinate $m$; $M=3a$ is unit cell length of the armchair ribbon.\cite{Xu08, corrugation09, zigzag_localization} Having calculated the Bloch states and constructed the band diagram one can readily obtain the number of the Bloch states $N^{Bloch}$ for a given Fermi energy that in turn serves as a basis for analysis of transport properties of GNCs.\cite{corrugation09} Because the projections of the two Dirac points in the armchair ribbon coincide at $k=0$ we solve \eqref{eq:Schrodinger} at zero wave vector and find $V_c$ from eigenvalues $E(k=0)$. 

 The integration path in \eqref{density} goes along the real axis and a fine integration grid is used to capture the locations of the subband edges and  quasibound states if any are present.
 
Since the Hartree potential $V_H$ given by Eq. \eqref{eq:V_H} depends on the electron density $n(\mathbf{r})$ which is a solution of the Schr\"{o}dinger equation with the Hamiltonian \eqref{eq:hamiltonian}, these equations need to be solved iteratively. The iteration process is executed until the convergence criterion $\frac{n^m_{out}-n^m_{in}}{n^m_{out}+n^m_{in}} < 10^{-3}$ is met, where $n^m_{in}$ and $n^m_{out}$ are the input and output average values of the electron density at the $m$-th iteration. In the cases where a constriction is present in the GN the above computation proceeds in two stages. At
the first stage, self-consistency is achieved for the uniform ribbon. Then constriction is imposed on the ribbon and self-consistency is achieved
again. 

Having calculated the electron density and the position of the Dirac point numerically, we are in a position to find the conductance 
\begin{equation}
	G = -\frac{2e^{2}}{h} \int dE \; \sum_{ij} T_{ij}(E) \frac{\partial f(E-E_F)}{\partial E}
	\label{eq:conductance}
\end{equation}
as a function of the Fermi energy. Here $T_{ij}(E)$ is the transmission coefficient from subband $j$ in the left lead to the subband $i$ in the right lead, at energy $E$. $T_{ij}(E)$ is calculated by the recursive Green's function method, see Ref. \onlinecite{Xu08} for details.

The Fermi energy and charge neutrality point are related to a value of the gate voltage measured in experimental setup as $V_c+E_F=eV_g$.\cite{FR07, Shylau09}  In a real device, it is the gate voltage $V_g$ that results in a change of carrier density in the graphene ribbon. We define $eV_g$ as the chemical potential difference between the metallic gate and the ribbon, necessary to accommodate extra carriers in the graphene and remove them from the metallic gate. Knowledge of both $V_c$ and $E_F$ thus allows one to estimate the value of $V_g$ used in an experiment. Note that $V_c=0$ in the non-interacting approach.

\section{Results}
\label{Results}

\begin{figure}[b!]
\includegraphics[scale=1.0]{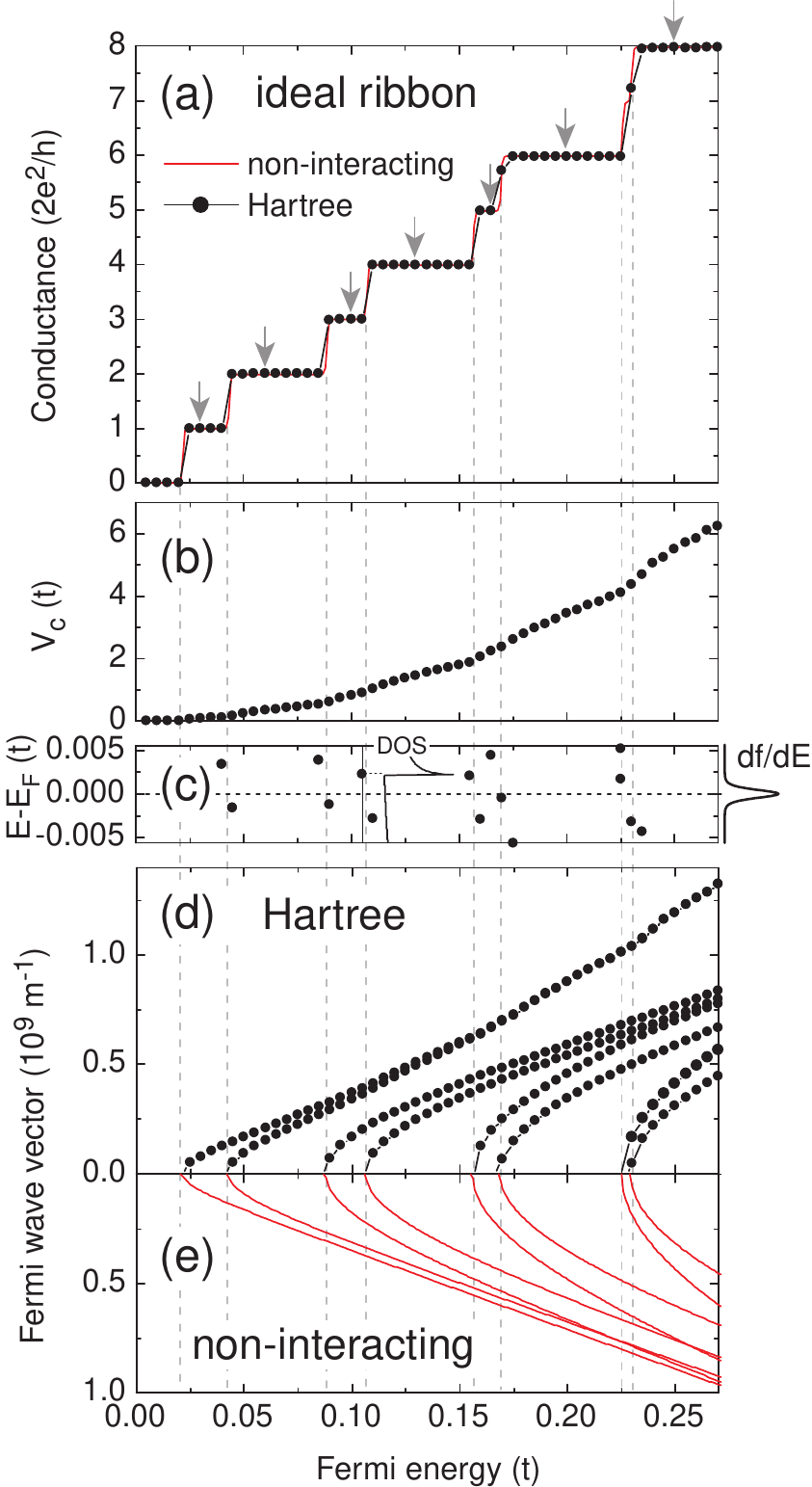}
\caption{(color online) Results for uniform GNs: The conductance (a), charge neutrality point (b), positions of the bottoms of the subbands that are near the Fermi energy (c), and the wavevectors at the Fermi energy calculated within the non-interacting (d) and Hartree (e) approaches vs. the Fermi energy. The inset in (c) shows the DOS for $E_F=0.105t$; the peak marks the position of the bottom of the subband. The outset in (c) shows the derivative of the Fermi-Dirac distribution function for $T=10$ K. Arrows in (a) mark the energies used for charge density plots in Fig. \ref{fig:3}. $t=2.7$ eV.}
\label{fig:2}
\end{figure}

\begin{figure}[th]
\includegraphics[keepaspectratio,width=\columnwidth]{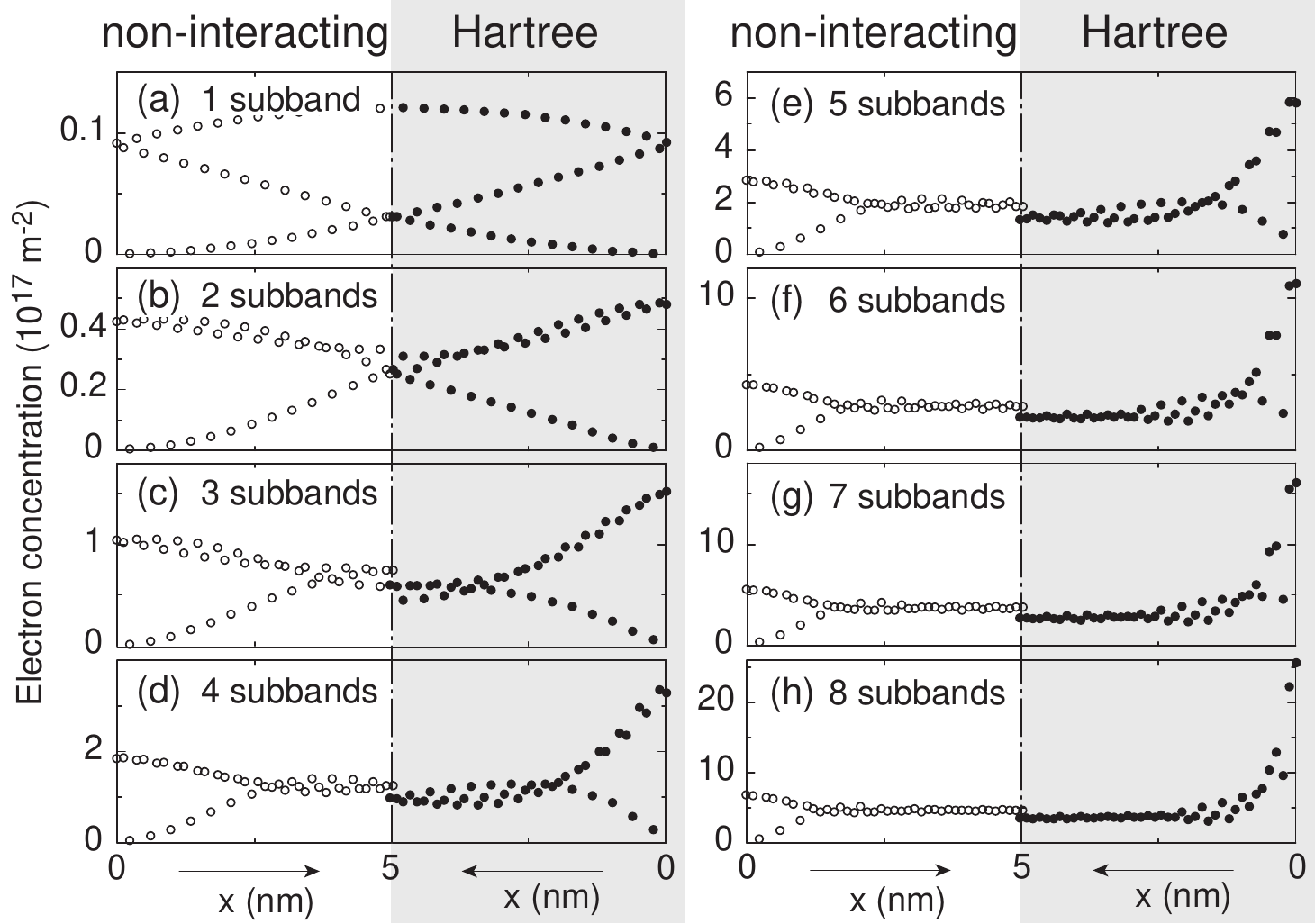}
\caption{Comparison of the electron concentrations in uniform GNs calculated in the non-interacting and Hartree approaches for different numbers of populated subbands. The electron concentration in the armchair unit cell oscillates between neighboring carbon atoms due to the specific structure of the wave functions, see e.g. Ref. \onlinecite{Brey06}. A half of the cross-section is shown. The electron Fermi energies for the different plots are indicated by arrows in Fig. \ref{fig:1}(a).}
\label{fig:3}
\end{figure}

Figure 2 shows results for uniform GNs: (a) the conductance, (b) the charge neutrality point, (c) the positions of the bottoms of the subbands that are near the Fermi energy, and (d) and (e) the wavevectors at the Fermi energy calculated within the Hartree and non-interacting models, respectively. The conductance values for the two models for the same values of the Fermi energy are very close to each other. In both cases the conductance increases by the quantum $2e^2/h$ each time a new subband opens for propagation. The opening of a new electron subband also results in an increase in the slope of the charge neutrality point $V_c$ vs. Fermi energy as is seen in Fig. \ref{fig:2}(b);  $V_c$ increases monotonically with increasing Fermi energy because the electron density on the ribbon increases. The slope change is caused by the additional contribution to the electrostatic potential on the ribbon due to the charges populating a new subband. Assuming a parabolic dispersion near the subband edge yields an $E^{-1/2}$ divergence  of the DOS that leads to an additional electron density $\delta n \propto \int E^{-1/2} dE \propto E^{1/2}$ where E is measured from the subband edge. The associated charge contributes to the Hartree potential that in turn leads to the rise in $V_c$. 

The subband energy position shown in Fig. 2(c) reveals a linear drop of the energy levels relative to $E_F$ as $E_F$ increases. Each time an energy level crosses $E_F$ electrons start populating the GN and contribute to the electrical conductance. Note that the bottoms of the subbands in the GN do not show any pinning to the Fermi level such as that observed in conventional quantum wires\cite{Hirose01} and open quantum dots\cite{opendot}. As can be seen in Fig. \ref{fig:2}(d), the electron interactions modify the band structure of the GN: The results for the Hartree model show avoided crossings and the two
lowest subbands having smaller velocities. The reason for these modifications of the band structure due to electron-electron interactions can be understood from the analysis of the charge distribution in the GN that is shown in Fig. \ref{fig:3}: The electron interactions in the Hartree model result in strong redistribution of the charges towards the edges of the ribbon when the Fermi energy is increased. The larger the Fermi energy, the stronger redistribution of the electron density. Note that the charge accumulation along the boundaries of uniform graphene strips was also discussed in Refs. \onlinecite{FR07, Silvestrov08, Shylau09, others_ee}. 

\begin{figure}[t]
\includegraphics[keepaspectratio,width=\columnwidth]{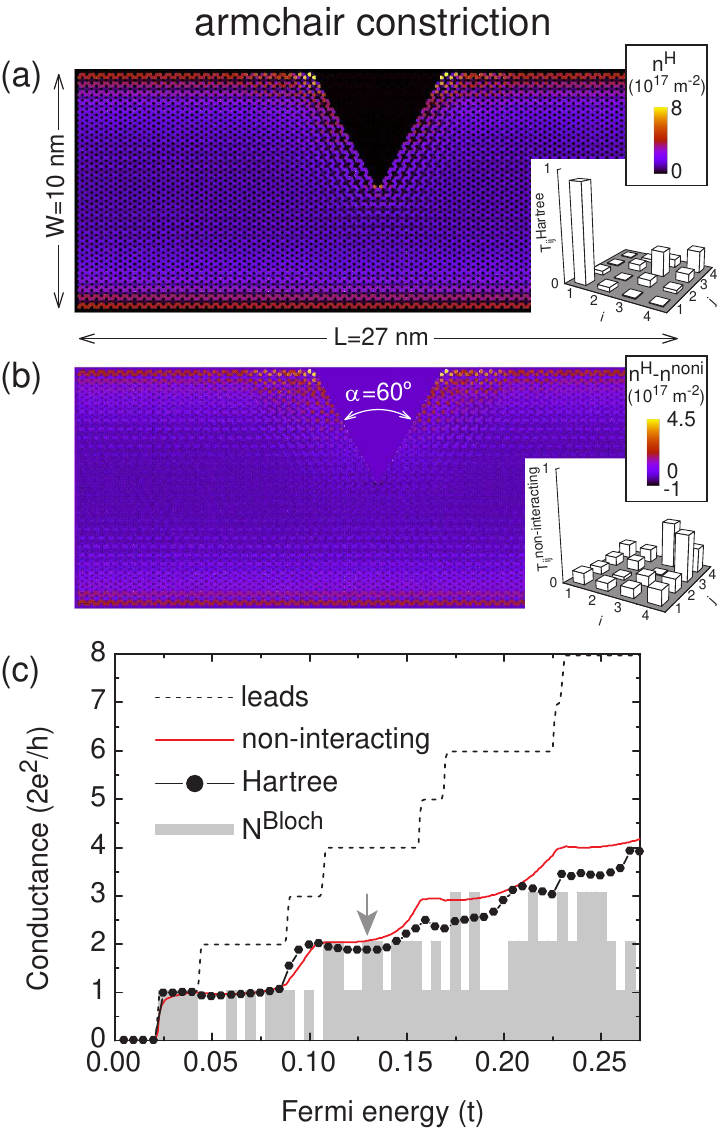}
\caption{(color online) (c) The conductance as a function of the Fermi energy for GN with armchair constriction: solid line - non-interacting electrons; line with dots - Hartree model, dashed line - (non-interacting) conductance without constriction that equals the number propagating states in the leads. The shaded gray area denotes the number, $N^{Bloch}$ of propagating Bloch states at the Fermi energy for the infinite modulated ribbon in which the constriction is repeated periodically. (a) The electron concentration $n^H$ calculated in the Hartree model. (b) The difference between Hartree and non-interacting electron concentrations $n^H-n^{noni}$. The Fermi energy in (a) and (b) is $E_F=0.13t$ and is marked by the arrow in (c). The insets in (a) and (b) show the partial transmission probabilities $T_{ij}$ for the Hartree and non-interacting calculations, respectively. }
\label{fig:4}
\end{figure}

\begin{figure}[thb]
\includegraphics[keepaspectratio,width=\columnwidth]{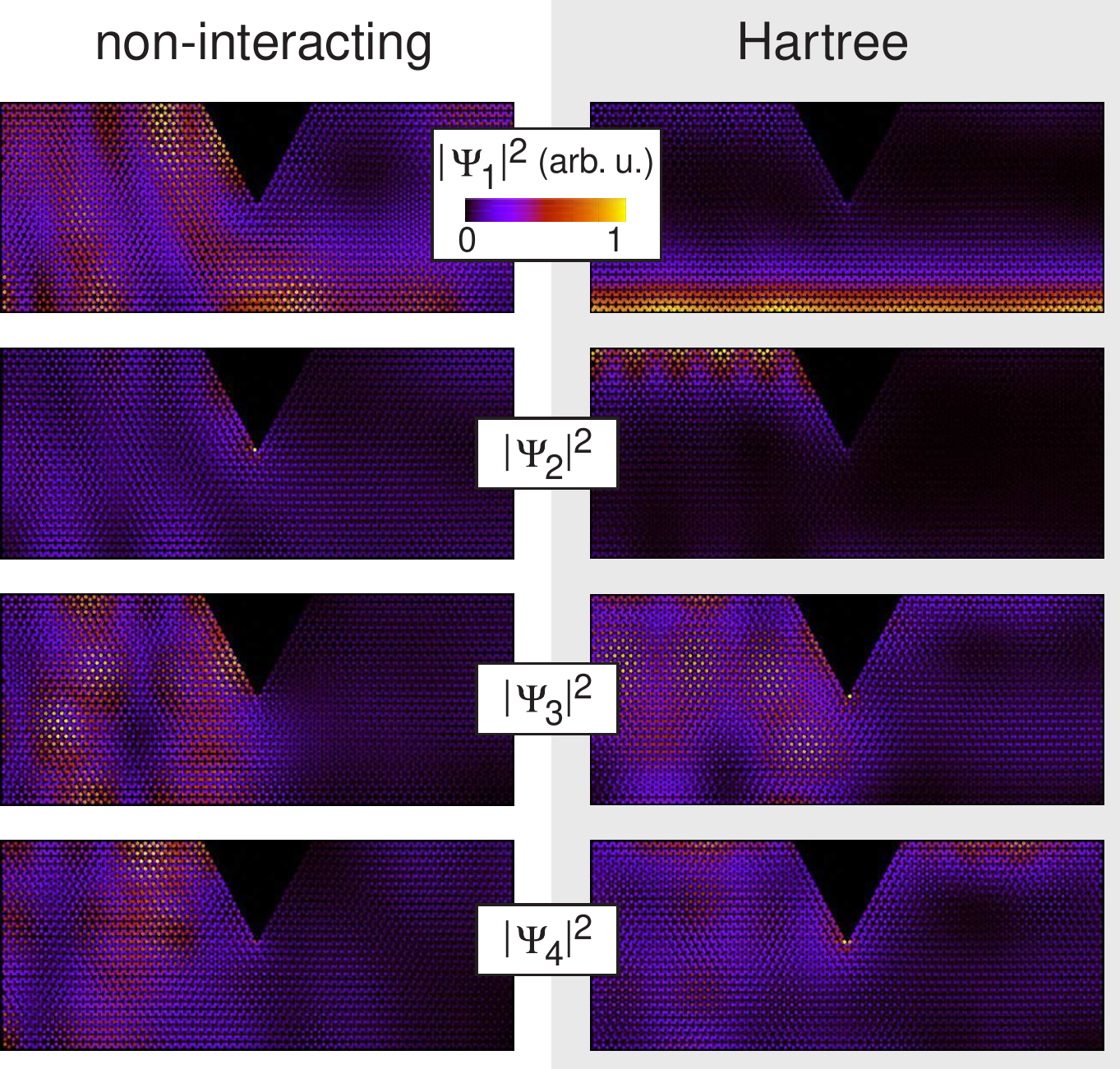}
\caption{(color online) The wave function square modulus $|\Psi_i|^2$ for a GN with an armchair constriction calculated within the non-interacting and Hartree approximations, left and right columns, respectively. The Fermi energy is $E_F=0.13t$, see the arrow in Fig. \ref{fig:4}(c). The corresponding partial transmission probabilities for the $i$th-state are shown in the insets in Fig. \ref{fig:4}(a),(b).}
\label{fig:7}
\end{figure}

Figure \ref{fig:4}(c) shows the conductance as a function of the Fermi energy for a GN with an armchair constriction. The non-interacting electron calculation predicts conductance quantization in steps of $2e^2/h$ over the whole range of $E_F$ values shown in Fig. \ref{fig:4}(c). However, the results of the Hartree calculation show only first two conductance plateaus to be well defined. The transition between these plateaus appears sharper than that predicted by the non-interacting calculation. In order to understand the reason for the better quantization for the first few conductance steps in the Hartree approach let's inspect the electron concentration distributions and transmission coefficients for $E_F=0.13t$ shown in Fig. \ref{fig:4} (a) and (b). The charge density is enhanced along the bottom straight boundary in Fig. \ref{fig:4} (a) and (b) in a similar way to that in the uniform GN discussed above. However the enhancement that occurs along the armchair constriction is very non-uniform: Much less charge accumulates near the apex at the narrowest part of the constriction; see Figs. \ref{fig:4}(a) and (b), respectively, for the electron concentration distribution $n^{H}(\mathbf{r})$ in the Hatree calculation and the difference between the Hartree and non-interacting electron concentrations $n^{H}(\mathbf{r})-n^{noni}(\mathbf{r})$. The partial transmission $T_{ij}$ in the Hartree model reveals nearly perfect transmittance for the first state, see inset in Fig. \ref{fig:4}(a). Note that $T_{11}\sim 1$ for all of the conductance steps in the Hartree theory. However, the transmittance due to the second state was found to be strongly suppressed. $T_{ij}$ for the non-interacting approach shows a fairly uniform distribution over all states, see inset in Fig. \ref{fig:4}(b). To understand this phenomenon, let's consider the square moduli of the wave functions shown in Fig. \ref{fig:7}. The first state in the Hartree approach transmits nearly perfectly because of its localization near the straight bottom boundary. By contrast, the second state is mostly localized near the top boundary, where the constriction is located and transmission is therefore blocked. Note that localization develops gradually as $E_F$ increases and the first and second states propagating along opposite boundaries reveal very similar dispersion for the uniform ribbon for $E_F\geq0.15t$, see Fig. \ref{fig:2}(d). These two states become mostly trapped within triangular wells at the straight ribbon boundaries. It is also worth noting that this phenomenon holds true for all of the constriction shapes studied below. It does not occur in the non-interacting electron model because of the absence of change accumulation near the boundaries. In general, we find electron interactions to favor intra-subband scattering whereas in the non-interacting approximation inter-subband scattering predominates. The conductance step degradation in the Hartree model for $E_F > 0.2t$ in Fig. \ref{fig:4}(c) is related to the overall poor transmittance of the highest states, where the charge accumulation along the ribbon's edges but not in the constriction itself is further increased. When this happens the constriction becomes a more effective obstacle to electron propagation through the ribbon in the Hartree theory than in the non-interacting electron approximation. 

\begin{figure}[t]
\includegraphics[keepaspectratio,width=\columnwidth]{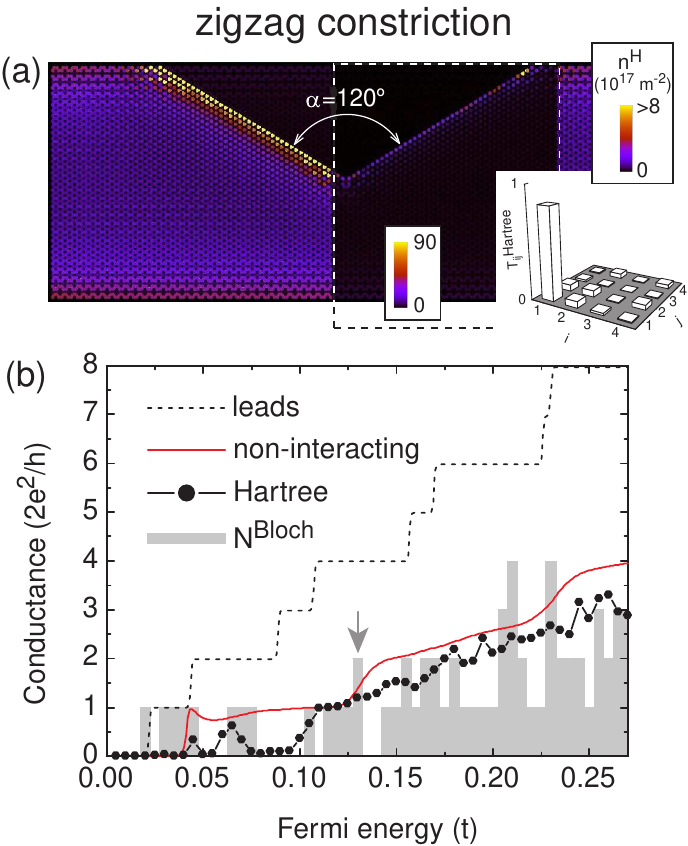}
\caption{(color online) (a) The electron concentration $n^H$ in the Hartree approximation. Concentrations equal to or greater than 8$\times10^{17}$ m$^{-2}$ are colored yellow. The dashed rectangle bounds region with the electron concentration plotted on a different scale. The inset in (a) shows the partial transmission probabilities $T_{ij}$ in the Hartree model. (b) The conductance as a function of the Fermi energy for a GN with a zigzag constriction. The labels and meaning of the shading are the same as in Fig. \ref{fig:4}(c). }
\label{fig:5}
\end{figure}

The conductance as a function of $E_F$ and a representative electron concentration distribution $n^H$ for a zigzag constriction are shown in Figures \ref{fig:5}(b) and (a). The charge density along the zigzag constriction edge is strongly enhanced by values up to an order of magnitude larger than for the armchair constriction; see the density scales in the inset in Fig. \ref{fig:5}(a). One reason for this is the electron localization at zigzag edges previously predicted in both non-interacting and interacting electron theories; see, for example, Refs. \onlinecite{corrugation09, zigzag_localization, Brey06}. This is a topological property of zigzag-terminated ribbons. Another reason is the effect of electron interactions that increases the charge density along the edge further. The electrons occupy only one graphene sublattice along the zigzag edge while they occupy both sublattices along the armchair edge. We find the crossover between these charge occupations to occur over a distance of about 10 carbon atoms at the armchair-to-zigzag junction. The partial transmission for the first subband is $T_{11}\sim 1$ similarly to the case of the armchair constriction although it is suppressed at $E_F\sim0.05t$ and $\sim0.08t$ due to resonant backscattering by strongly localized states at the zigzag edge. It is worth noting that in modulated ribbons consisting of periodically repeated identical constrictions stop-bands form at these energies. These stop-bands, where the number of propagating Bloch states $N^{Bloch}$ (the grey shading in Fig. \ref{fig:5}(b)) in the periodic structure is zero, are similar to those predicted in edge-corrugated graphene ribbons.\cite{corrugation09} However, here the stop-bands form due to electron interactions. The mismatch between the electronic structures of the armchair host and zigzag constriction also contributes to less pronounced conductance quantization being seen in Fig. \ref{fig:5}(b) for the zigzag constriction than in Fig. \ref{fig:4}(c) for the armchair constriction. 

\begin{figure}[t]
\includegraphics[keepaspectratio,width=\columnwidth]{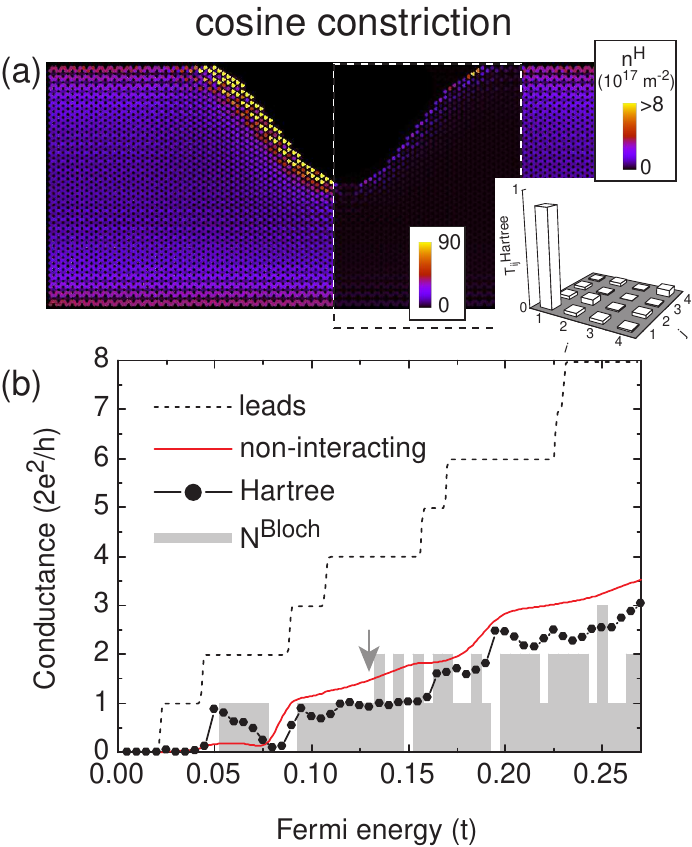}
\caption{(color online) The same as Fig. \ref{fig:5} but for a GN with a constriction having a cosine profile.}
\label{fig:6}
\end{figure}
  
Figure \ref{fig:6} shows a representative electron concentration distribution $n^H$
and the conductance for a constriction with a cosine profile.
We found strong accumulation of the charge density along the
boundary of the constriction itself, similar to that for the zigzag constriction.
Note that even though the
overall shape of the cosine constriction is smooth there are multiple
pieces of the zigzag terminated edges connected to each other by atomic
steps along the constriction boundary. The conductance dip at $E_F\sim0.08t$ is correlated to the absence of propagating Bloch states at this energy in the corresponding modulated ribbon with periodically repeated constrictions, as is the case for the similar features in Fig. 6(b) for the zigzag constriction. Interestingly, although in Fig. \ref{fig:6}(b)
only hints of conductance plateaus can be seen in the results for the non-interacting
electron model (red solid curve), a very pronounced first conductance plateau is found
when electron interactions are included in the Hartree model (dots), although only hints
of higher plateaus can be discerned in this case as well. We find the first conductance plateau to remain robust for GNCs with the cosine shape of length 7-16 nm (not shown). It may be relevant that in their experimental
study of a constriction with an overall smooth profile (albeit for a much larger structure
than those studied here) Tombros {\em et al.}\cite{Tombros11} observed a very pronounced first
integer conductance plateau $(g \sim 2e^2/h)$ but only very weak higher plateaus or only
hints of higher plateaus. In such a scenario, a well defined first plateau may be caused by the lowest state being adiabatically transmitted along one of the graphene boundaries in an asymmetric constriction with curvature of one boundary being much smaller than that of the other. Because no image of the constriction on which the transport measurements were carried out was presented\cite{Tombros11}, the degree of asymmetry is not known. Note also that the experimental device in Ref. \onlinecite{Tombros11} was fabricated by current annealing that substantially reduces edge disorder at both boundaries.

\section{Summary}
\label{Summary}
In conclusion, we have presented a self-consistent model of 
electron quantum transport in graphene ribbons and constrictions. The latter are represented as trenches of depth 5 nm and length range 7-17 nm and having different shapes. The model is based
on the Green's function formalism and accounts for electron-electron
interaction within the Hartree approach. 
The Hartree model predicts several
features not found in the non-interacting model. The electron charge density gradually
accumulates along straight boundaries of uniform ribbons as the Fermi energy increases. However, accumulation at the constriction depends strongly on the details of the constriction geometry. There is little if any charge accumulation 
along a V-shaped armchair constriction boundary but strong accumulation along 
a V-shaped zigzag boundary or a boundary with an overall smooth cosine profile.
For each of these constriction types imposed on a ribbon with armchair boundaries,
except near isolated reflection resonances, we find almost perfect transmission of electrons in the first 
subband (with little inter-subband scattering)  and almost perfect reflection of electrons in the second 
subband of the ribbon in the Hartree model. By contrast, we find the constriction 
to induce strong intersubband scattering of electrons for every subband in the non-interacting electron model. 
For the constriction with the cosine profile, the first integer conductance plateau 
is much more pronounced in the Hartree model than in the non-interacting model,
a finding that may be relevant to the recent experiment of Tombros {\em et al.}\cite{Tombros11}.
The transport properties of two lowest subbands are the result of by electron localization near the opposite boundaries of the ribbon. An analogy can be drawn between perfect transmission along a uniform graphene boundary and the edge state transport in the quantum Hall effect, where both are immune to defects in interior of a device.

\begin{acknowledgments}
This research was supported by CIFAR, NSERC, Compute Canada and Westgrid. 
\end{acknowledgments}

\end{document}